\documentclass[12pt,preprint,tighten]{aastex}
\newcommand{\pref}{\protect\ref}
\begin{document}

\shorttitle{SUMER Coronal Hole Partitioning}
\shortauthors{S.~W. McIntosh et~al.}
\title{Simple Magnetic Flux Balance as an Indicator of Neon VIII Doppler Velocity Partitioning in an Equatorial Coronal Hole}
\author{Scott W. McIntosh}
\affil{Southwest Research Institute, Department of Space Studies,\\
1050 Walnut St, Suite 400, Boulder, CO 80302}
\email{mcintosh@boulder.swri.edu}
\author{Alisdair R. Davey}
\affil{Southwest Research Institute, Department of Space Studies,\\
1050 Walnut St, Suite 400, Boulder, CO 80302}
\email{ard@boulder.swri.edu}
\and
\author{Donald M. Hassler}
\affil{Southwest Research Institute, Department of Space Studies,\\
1050 Walnut St, Suite 400, Boulder, CO 80302}
\email{hassler@boulder.swri.edu}

\begin{abstract}
We present a novel investigation into the relationship between simple estimates of magnetic flux balance and the \ion{Ne}{8} Doppler velocity partitioning of a large equatorial coronal hole observed by the Solar Ultraviolet Measurements of Emitted Radiation spectrometer (SUMER) on the {\em Solar and Heliospheric Observatory} (SOHO) in November 1999. We demonstrate that a considerable fraction of the large scale Doppler velocity pattern in the coronal hole can be qualitatively described by simple measures of the local magnetic field conditions, i.e., the relative unbalance of magnetic polarities and the radial distance required to balance local flux concentrations with those of opposite polarity.
\end{abstract}

\keywords{Sun: transition region \-- Sun:magnetic fields \-- Sun: granulation \-- Sun: solar wind \-- Sun: UV Radiation}

\section{Introduction}

The Solar Ultraviolet Measurements of Emitted Radiation spectrometer \citep[SUMER;][]{Wilhelm+others1995} on the ESA/NASA {\em Solar and Heliospheric Observatory} \citep[{\em SOHO};][]{Fleck+others1995} has been observing the Sun with high spectral and spatial resolution since 1996 January. A considerable number of SUMER observations are in the form of long-duration slit-scan ``rasters'' (spectro-heliograms) of polar and equatorial coronal hole regions. These coronal hole rasters have been used to investigate the spatial correspondence between supergranular network cells, visible in the photosphere and chromosphere, and their counterpart emission and plasma flows observed in the solar transition region and lower corona. The analysis of these SUMER rasters has placed particular emphasis on the study of the \ion{Ne}{8} 770~\AA{} emission line, identifying and investigating regions of significant plasma outflow ($\ge$5~km~s$^{-1}$) in the 600,000~K plasma as a physical link to the origin of solar wind outflow \citep[e.g.,][]{Warren+others1997,Hassler+others1999,Dammasch+others1999,Tu+others1999,Tu+others2005a,Tu+others2005b, Xia+others2003,Xia+others2004,Wiegelmann05,Aiouaz2005}.

In this Letter, we discuss and demonstrate two new and very simple diagnostics of the magnetic environment encompassing a specific set of SUMER coronal hole rasters: the relative unbalance of magnetic polarities on the supergranule spatial scale, and the radial distance required to balance local flux concentrations with those of opposite polarity. We demonstrate that the partitioning (i.e., how the Doppler image can be broken into regions of blue- and red-shift) of the \ion{Ne}{8} 770~\AA{} Doppler shifts and emission in the upper transition region of the coronal hole can be qualitatively understood by imposing a simple model of magnetic flux balance. Further, we observe that the magnitude of the net magnetic unbalance in the coronal hole appears to have a critical influence on the Doppler shift patterning, and on the blue-shift in particular.

\section{Observations and Data Reduction}\label{sec:data}
We focus our discussion on two individual SUMER coronal hole rasters\footnote{Each of the rasters was constructed using SUMER's Detector A, the 1\arcsec x 360\arcsec{} (Slit~2) arrangement and a 3\arcsec{} raster step size in the (solar) West to East direction with an exposure time of 150s. } from 1999 November 6 and 7 that, when combined, form a synoptic scan of a large equatorial coronal hole (ECH) in the 1530 to 1555~\AA{} range. This spectral range includes four strong emission lines with formation temperatures representative of the upper chromosphere (\ion{Si}{2}; 1533.4\AA \-- 30,000~K), transition region (\ion{C}{4}; 1548.2 and 1550.8~\AA \-- 100,000~K) and the upper transition region/low solar corona (\ion{Ne}{8}; 1540.8~\AA{} in second order \-- 770.4~\AA{} in first order \-- 600,000~K). In this Letter, we concentrate only on the Doppler patterning of the \ion{Ne}{8} emission line and its connection to the observed global morphology of the ECH region. 

A recent paper \citep{Davey2006} has developed an improved method to correct small (persistent) electronic imperfections of SUMER's detector A, which can cause systematic shifting of the resulting solar UV spectra when acquired in raster mode. By fitting the Gaussian profiles of ten accurately measured (in the laboratory) \ion{Si}{1} emission lines in the 1530-1555~\AA{} range at each spatial row of detector pixels and mapping their locations onto the plane of the detector, they are able to build a map of the detector distortion that is caused by the small electronic defects. Using this corrective offset to ``co-align'' the \ion{Si}{1} emission lines on each detector image in the raster, they are able to compose a self-consistent wavelength calibration for the entire data set. The net result of this wavelength calibration is a set of Doppler velocities in the \ion{Ne}{8} emission line to an accuracy of about $\pm$1~km/s. Further,  following the discussion presented in Sect.~4.1 of \citet{Davey2006}, we have chosen to use a rest wavelength of 770.420\AA{} for this line, consistent with the laboratory measurement of \citet{Fawcett1961}.

%\placetable{tab:one}
%\placefigure{fig1}

Details of the SUMER raster observations of November 6 and 7 are summarized in Table~\pref{tab:one} and are pictorially represented in the panels of Fig.~\pref{fig1}. As discussed in \citet{Davey2006}, we have rotated\footnote{All translation, rotation and co-alignment operations are performed on the {\em SOHO} data maps using Dominic Zarro's mapping routines that can be found at - http://orpheus.nascom.nasa.gov/$\sim$zarro/idl/maps.html .} both raster ``maps'' to their coordinates at the start time of the November 6 raster, as this facilitates co-alignment with the other {\em SOHO} instruments.

The left panel of Fig.~\pref{fig1} shows the {\em SOHO} Extreme-ultraviolet Imaging Telescope \citep[EIT;][]{DelaboudiniereEA95_short} 195 \AA{} passband coronal context image from 1999 November 6 taken closest to the start of the first raster. Each EIT 195\AA{} passband image contains broadband spectral (multi-thermal) emission that is strongly peaked at a little over 1.5~MK \citep[see, e.g., Figs. 8b and 9b of][]{DelaboudiniereEA95_short} from \ion{Fe}{12}. The right panel of Fig.~\pref{fig1} shows the {\em SOHO} Michelson Doppler Imager \citep[MDI;][]{Scherrer1995} line-of-sight magnetogram taken at 12:47~UTC, rotated to the time of the EIT image at 14:05~UTC. As well as the SUMER raster regions in the panels we show the (estimated) boundary of the ECH, which we will set at an EIT iso-intensity contour of 150 Data Numbers (DN). We note that the global morphology of the corona hole region does not change over the twenty hours of observation required to build the two SUMER rasters and so the step of comparing them to the context coronal images, implied ECH boundary, and photospheric magnetogram at the same reference time is assumed to be valid.

Figure~\pref{fig2} shows the coronal hole region of 1999 November 6 as determined by a numerical adaptation of the Harvey method for coronal hole identification \citep[][]{Harvey2002}. The automated coronal hole identification tool \citep{Henney2005}, like its counterpart, uses the equivalent width of \ion{He}{1} 10830~\AA{} line determined from daily synoptic Kitt Peak Vacuum Telescope (KPVT) spectroheliograms and magnetograms to morphologically isolate regions of reduced equivalent width. The red- or blue-colored regions indicate the net magnetic polarity of the KPVT coronal hole regions identified by the code as positive or negative respectively. We note that the three-pronged EIT 195~\AA{} contour (centered on $-$200\arcsec{}, $-$400\arcsec{}) that is clearly not identified as part of the coronal hole from the KPVT boundary is a filament channel and is quite distinct from the coronal hole in other EUV wavelengths (J.B. Gurman \-- private communication). Other than the filament channel, we see that the heuristic EIT coronal hole boundary and that numerically determined from KPVT observations agree well. We will therefore continue to use the simple EIT boundary of the ECH in the discussion that follows without any loss of generality.

%\placefigure{fig2}

The top row of panels in Fig.~\pref{fig3} show the \ion{Ne}{8} intensity (left) and Doppler velocity (right) maps from the regions rastered by SUMER on November 6 (upper) and November 7 (lower), respectively. Again, we show the EIT 195~\AA{} 150~DN iso-intensity contour for reference as the ECH boundary. We notice that while the \ion{Ne}{8} line intensity raster map from November 6 (North-East portion) is almost uniformly dark in the coronal hole, with emission rising at the boundary, the November 7 raster map shows a ``break'' in the emission structure (centered on 100\arcsec{},$-$250\arcsec{} \-- running from seven to two o'clock), where the intensity increases by a factor of two, to the quiet-Sun level (i.e., that outside the hole boundary) and drops again in the lower right (South-West) portion of the coronal hole. Very close inspection of the EIT image shows some (very low) enhanced emission contrast in the break region of the ECH, but it does not reflect the factor of two change in the \ion{Ne}{8} emission.

%\placefigure{fig3}

Precisely the same spatial break, or partition, exists in the Doppler shift rasters shown in the top right panel of the figure. The raster from November 6 shows a large, complex pattern of \ion{Ne}{8} blue-shifts inside the EIT contour, while that of November 7 shows a more divided picture with the break region showing a mottled, but net-postive (red), Doppler patterning that is reminiscent of the \ion{Ne}{8} quiet-Sun data shown to the East of the EIT contour on the sixth and those published in the literature \citep[e.g.,][]{Hassler+others1999,Dammasch+others1999,Dammasch2002,Wilhelm2002,Davey2006}. In the South-West portion of the November 7 raster  the Doppler-shift pattern again resembles the dark blue-shift structure of the North-Eastern coronal hole region of the previous day. \citet{Wilhelm2002} queried the strange emission and Doppler velocity partitioning of the November 7 1999 raster within the uniformly dark EUV emission that forms the body of the coronal hole. It is the (qualitative) understanding of the partitioning of the \ion{Ne}{8} emission and velocity structure and their connection to the global morphology of the ECH that provides the motivation for this Letter. 

\subsection{Magnetic Flux Balance Diagnostics}
The bottom row of panels in Fig.~\pref{fig3} show the 20~Mm radially averaged line-of-sight magnetic field ($\langle B_{||} \rangle_{r=20}$; left) and the ``Magnetic Range of Influence'' (MRoI; right). The first is a relatively simple diagnostic which highlights the net magnetic field variation at the spatial scale of supergranular cells \citep[15\--30Mm;][]{Hagenaar1997}. It shows that the bulk of the region in the North-East of the coronal hole has a dominant and non-negligible net positive magnetic flux, as does the strongly blue-shifted region in the South-West of the November 7 raster, while the central ``break'' region is near-zero, mixed magnetic polarity and threaded by the neutral line, or where $\langle B_{||} \rangle_{r=20}$ is identically zero. The net positive magnetic polarity of the entire coronal hole can be easily compared to the KPVT coronal hole boundary determination in Fig.~\pref{fig2}. 

The MRoI is another simple realization of the magnetic environment and reflects the (radial) distance required to balance the integrated magnetic field contained in a 25~Mm$^2$ circular area centered at any pixel in the MDI magnetogram. It is computed by advancing radially outward from the central region one pixel at a time and summing the contained field to seek the distance (the number of MDI pixels times their spatial scale, $\sim$1.9~Mm) where the total summed field returns to zero ($<$1~G). While the MRoI contains {\em no} directional information, it can be thought of as a measure of magnetic partitioning of the plasma into open and closed regions. When the MRoI is small, the magnetic field is ``closed'' locally, while, if the MRoI is large, the magnetic field at the center is largely unbalanced and the magnetic environment is effectively ``open''. 

\section{Topological Influence on the \ion{Ne}{8} 770\AA{} Rest Wavelength}
Considering again our choice of 770.420\AA{} as the \ion{Ne}{8} rest wavelength, we point to the line intensity pattern of both rasters and its connection to the global magnetic environment of the region surveyed. While we know that the integrated line intensity is completely insensitive to the choice of rest wavelength, the observed Doppler patterning {\em must} reflect the variation of the intensity and magnetic environment because they are physically coupled. We have seen that the darkest regions of \ion{Ne}{8} emission are predominantly blue-shifted and correlate to regions of large MRoI and unbalanced magnetic fields. Conversely, the brighter regions are correlated to small MRoI and balanced magnetic fields. The spatially averaged Doppler velocity in the magnetically closed regions is slightly red-shifted \citep[1\--3~km/s][]{Davey2006}, which, when combined with the $\pm$1~km/s accuracy of the wavelength calibration, point to a rest wavelength of 770.420\AA{} and an error in the range of $\pm$1-4~m\AA. We argue that the results presented in this Letter add weight to our choice of 770.420\AA{} and are entirely consistent with the statistical justification of \citet{Davey2006}. We should declare that while this remotely sensed inference may be the most self-consistent measurement that we have from the solar plasma, it in no way replaces the need for new theoretical calculations or (technically difficult) controlled laboratory experiments to determine what is possibly the most physically important, and contentious, rest wavelength in the solar ultraviolet spectrum. A detailed investigation into the SUMER measurements of the \ion{Ne}{8} rest wavelength is beyond the scope of this Letter and will be revisited in a forthcoming paper (Davey et~al. 2006, in preparation).

\section{Discussion}\label{sec:discuss}
From the top row of panels of Fig.~\pref{fig3}, we are immediately drawn to three features: the locations of strong ($\ge$~5~km~s$^{-1}$) blue shift in the North-East and South-West of the EIT defined coronal hole region and the brighter \ion{Ne}{8} emission region in its central belt. While the quantitative comparison of the Doppler shift in the SUMER rasters with the $\langle B_{||} \rangle_{r=20}$ and MRoI  maps does not yield direct physical correlations, we can clearly see that the regions of strongest \ion{Ne}{8} blue-shift are co-spatial with regions of large-scale, locally unbalanced, net positive ($\ge$10~G) magnetic field. The mixed velocity \ion{Ne}{8} regions are those where the $\langle B_{||} \rangle_{r=20}$ and MRoI are both small, $\sim$0~G and $\le$ 20~Mm, respectively.

\begin{figure}
\figurenum{1}
\plotone{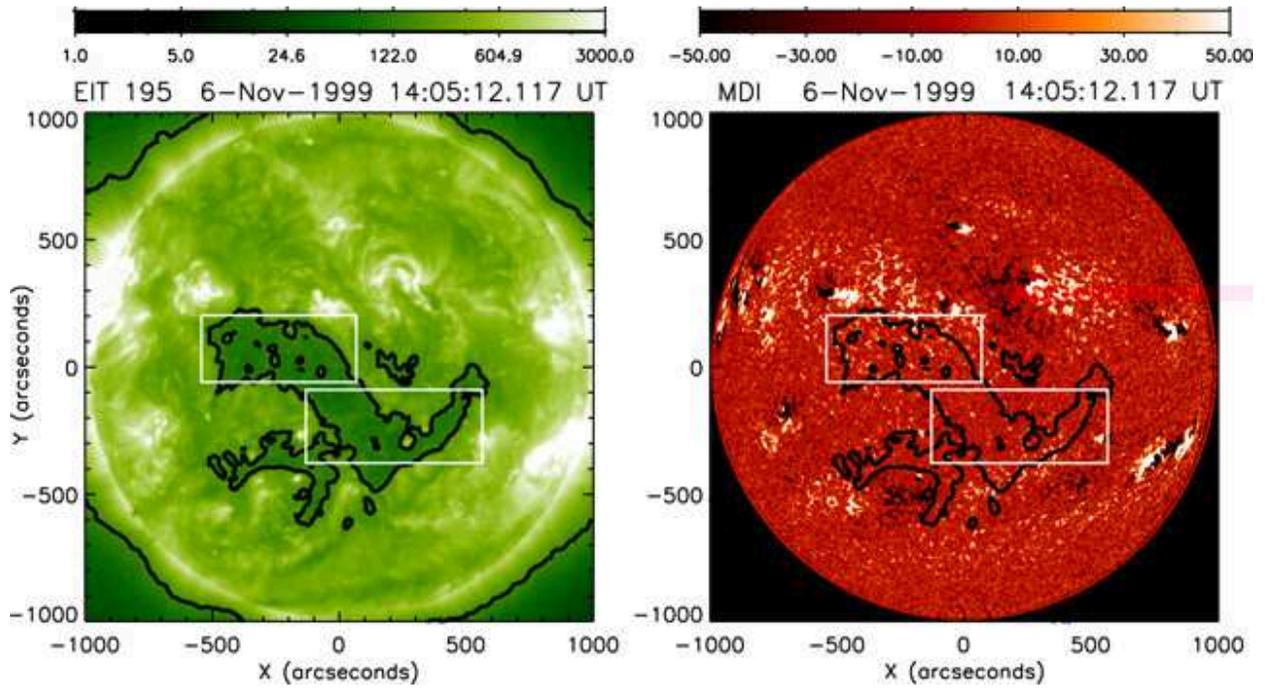}
\caption{The {\em SOHO} EIT 195\AA{} emission (left) and SOHO MDI line-of-sight magnetogram closest to the start of the SUMER raster on 1999 November 6. Each of the panels show the EIT~195\AA{} 150~DN contour in black to outline the coronal hole region and the regions rastered by SUMER on November 6 (top left) and November 7 (bottom right) as white rectangular outlines. \label{fig1}}
\end{figure}

\begin{figure}
\figurenum{2}
%\epsscale{0.65}
\plotone{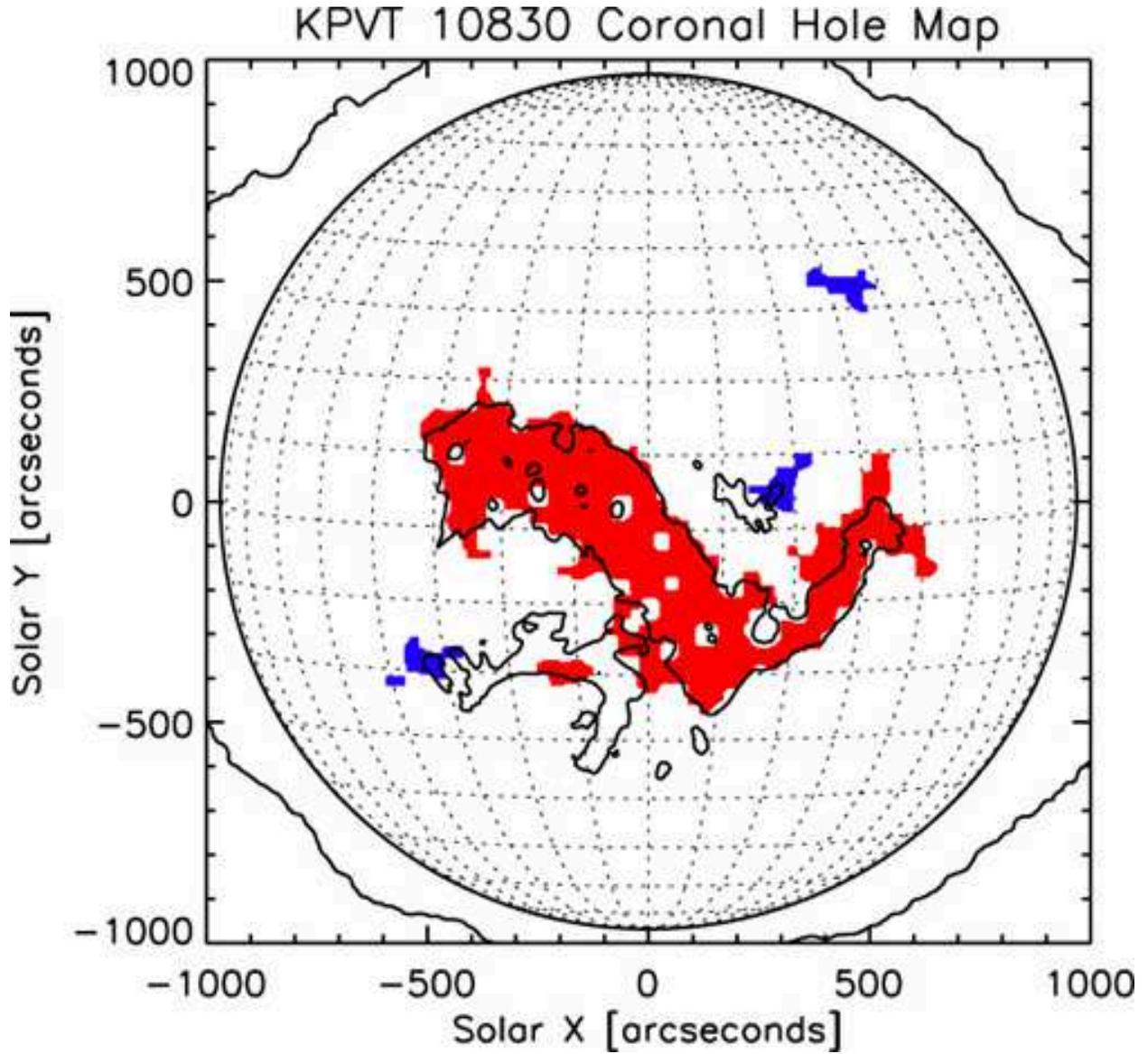}
\caption{The coronal hole boundary from 1999 November 6 determined from the equivalent width of KPVT \ion{He}{1} 10830~\AA{} emission (see text for details). We also show the EIT 195\AA{} 150~DN contour in black for comparison. \label{fig2}}
\end{figure}

\begin{figure}
\figurenum{3}
%\epsscale{0.85}
\plotone{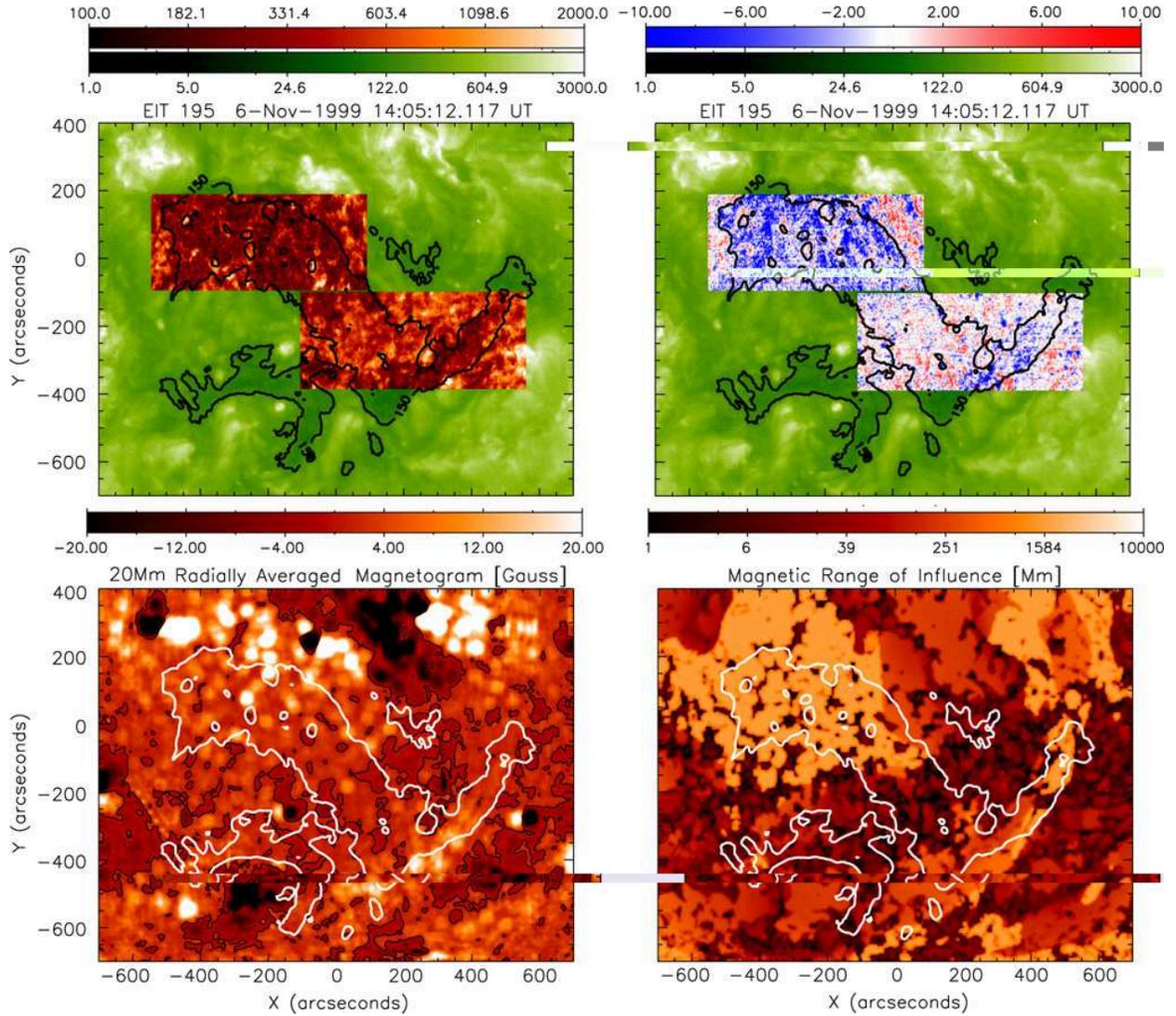}
\caption{From top to bottom, left to right, the SUMER \ion{Ne}{8} 770~\AA{} integrated line intensity (Units: Counts), Doppler-velocity (km~s$^{-1}$), the supergranular radially averaged magnetic field strength ($\langle B_{||} \rangle_{r=20}$; Gauss) and the ``Magnetic Range of Influence'' (MRoI; Mm). On each of the panels in the figure we show the EIT 195~\AA{} 150~DN contour to outline the coronal hole region. Additionally, in the lower left panel we show the thin black contour that designated the magnetic ``neutral line'', where $\langle B_{||} \rangle_{r=20} = 0$~G.  \label{fig3}}
\end{figure}

We notice that the portion of the coronal hole where the \ion{Ne}{8} emission is enhanced is co-spatial with the region where the MRoI is $\le$20~Mm. We also see that this region is where the magnetic field strength averaged over supergranular spatial scales is, on balance, zero and is threaded by the neutral line. This portion of the coronal hole displays Doppler velocity patterning that is (qualitatively) closest to that of the quiet Sun \citep[e.g., Fig.~12 of][]{Davey2006}, or at least to the raster portions that are outside the coronal hole. We should note that there is no obvious enhancement in the EIT emission in the same location (see Fig.~\pref{fig1}). While it might be magnetically open in the hot solar corona (the uniformly dark EIT emission), it would appear that the region is locally closed, at least to the formation height of \ion{Ne}{8} in the transition region. The large-scale complexity of the \ion{Ne}{8} emission and Doppler velocity structure points to a considerably more complex coronal hole boundary than might simply be implied considering {\em only} the EIT 195~\AA{} emission, or the KPVT boundary for that matter. This quandary does raise an important, interesting and timely question to the community: How do we observationally determine the boundary of a coronal hole in a robust manner? 

We have observed that the Doppler partitioning and simple diagnostics of the magnetic environment imply that there are transition region structures on supergranular length scales (or larger) that are topologically closed (small $\langle B_{||} \rangle_{r=20}$ and MRoI) or open (larger $\langle B_{||} \rangle_{r=20}$ and MRoI). In general, the closed regions show enhanced line emission while the open regions show decreased emission and strong blue-shifts. We have also observed that the emission structure in the 1.5~MK corona does not always reflect that of the 600,000~K transition region. This strange feature implies the existence of ``nesting'' in the different scales of magnetic topology, such that the lower region can be almost entirely closed, while that above is open. While space limits any explanation of how the apparent nesting of the plasma arises, we can suggest one simple situation that could produce such an effect: two distinct coronal holes are present (delimited by the NE and SW \ion{Ne}{8} Doppler-shift regions), with the magnetically closed region in the transition region underneath the place where the divergent magnetic field lines from each merge in the corona. However, we note that many models of varying complexity can satisfy the global aspects of these observations; discussion is therefore left to later work.

In summary, the apparently different plasma loading, heating and acceleration that occurs between the open and closed magnetic environments in the transition region (and possibly into the hotter corona) is a subject requiring a detailed investigation of the local and global magnetic topologies. As such, it cannot be dealt with in this short Letter. However, based on the observational evidence, we can speculate that the magnetic environment leads to a preferential energy balance in the transition region plasma; the magnetically closed regions are dominated by plasma heating (enhanced thermal emission) while that of the open regions would appear to be dominated by kinetic energy (enhanced plasma outflow). Further, we can infer that, if the blue shifts in open field regions are indeed the origins of solar wind outflow \citep[e.g.,][]{Hassler+others1999,Tu+others1999,Tu+others2005a,Tu+others2005b, Xia+others2003,Xia+others2004}, it is possible that the presence of the unbalanced magnetic fields enables the plasma to create, load mass, heat and sustain the lowest tributaries of the solar wind through a magneto-convection driven mechanism like ``magnetic exchange reconnection'' \citep[][]{Wang1998,Priest2002}. This is a subject of ongoing research by the authors (McIntosh et~al. 2006, in preparation). 

\acknowledgements 
We would like to thank the anonymous referee, Werner Curdt, Stein Vidar Haugan, John Mariska, Klaus Wilhelm and Meredith Wills-Davey for carefully reading, commenting on and enhancing the manuscript. Similarly, we would like to thank Carl Henney for discussing his method of automatic coronal hole boundary determination. This material is based upon work carried out at the Southwest Research Institute that is supported in part by the National Aeronautics and Space Administration under grants issued under the Living with a Star and Sun-Earth Connection Guest Investigator Programs. Specifically Grants NAG5-13450 and NAG5-11594 to DMH and NNG05GM75G to SWM. The SUMER project is financially supported by DLR, CNES, NASA and the ESA PRODEX Program (Swiss contribution). SUMER is part of SOHO, the {\em Solar and Heliospheric Observatory}, of ESA and NASA.

\clearpage

\begin{table}
\caption{Details of the SUMER coronal hole rasters under discussion. \label{tab:one}}
\begin{tabular}{cccccc} \tableline \tableline
Observation & Observation & \multicolumn{1}{c}{Duration} &
\multicolumn{1}{c}{Solar X} & \multicolumn{1}{c}{Solar X} & Solar Y \\
Start & End &\multicolumn{1}{c}{(h)} & \multicolumn{1}{c}{at Start (\arcsec)} &
\multicolumn{1}{c}{at End (\arcsec)} & (\arcsec) \\ \tableline
 06-Nov-99 14:00:13 &  07-Nov-99 00:15:50 &  10.260 &  -535 &   199 & 45\\
 07-Nov-99 00:35:23 &  07-Nov-99 10:16:18 &   9.682 &    -4 &   708 &  -249\\ \tableline
\end{tabular}
\end{table}

\clearpage

\clearpage

\clearpage

\end{document}